\documentclass[aps,pra,groupedaddress,floats,reprint]{revtex4-1}

\usepackage[english]{babel}

\usepackage{amsfonts,graphicx,soul,natbib,mathrsfs}

\usepackage[colorlinks=true]{hyperref}
\usepackage[dvipsnames]{xcolor}
\usepackage[normalem]{ulem}
\usepackage{epstopdf}
\usepackage{textcomp}
\usepackage{xcolor}
\usepackage{soul}
\usepackage{physics}
\usepackage{graphics}
\usepackage{comment}
\usepackage{amssymb}
\usepackage{tensor}
\usepackage{array}
\usepackage{wrapfig}
\usepackage{multirow}

\usepackage{graphicx,subcaption}

\usepackage{animate}

\usepackage[dvipsnames]{xcolor}

\newcommand{\expv}[1]{\langle #1 \rangle}

\usepackage{dcolumn}
\usepackage{bm}

\begin{document}

\title{Transmission of vortex electrons through a solenoid}

\author{G.\,K.~Sizykh}
\email{georgii.sizykh@metalab.ifmo.ru}
\affiliation{School of Physics and Engineering,
ITMO University, 197101 St. Petersburg, Russia}

\author{A.\,D.~Chaikovskaia}
\affiliation{School of Physics and Engineering,
ITMO University, 197101 St. Petersburg, Russia}

\author{D.\,V.~Grosman}
\affiliation{School of Physics and Engineering,
ITMO University, 197101 St. Petersburg, Russia}

\author{I.\,I.~Pavlov}
\affiliation{School of Physics and Engineering,
ITMO University, 197101 St. Petersburg, Russia}

\author{D.\,V.~Karlovets}
\email{dmitry.karlovets@metalab.ifmo.ru}
\affiliation{School of Physics and Engineering,
ITMO University, 197101 St. Petersburg, Russia}

\begin{abstract}
We argue that it is generally nonstationary Laguerre-Gaussian states (NSLG) rather than the Landau ones that appropriately describe electrons with orbital angular momentum both in their dynamics at the boundary between a solenoid and vacuum and inside the magnetic field. It is shown that the r.m.s. radius of the NSLG state oscillates in time and its period-averaged value can significantly exceed the r.m.s. radius of the stationary Landau state. We propose several experimental scenarios to probe this unconventional quantum dynamics in the magnetic fields typical for electron microscopes and particle accelerators.
\end{abstract}

\maketitle


\textbf{Introduction}. Manipulation of electrons with orbital angular momentum (OAM), dubbed twisted or vortex electrons \cite{Bliokh2017, Lloyd2017}, is a useful tool with great prospects of applications in electron microscopy, nanomaterials studies, particle physics, accelerator physics, and other fields \cite{Verbeeck2010, Idrobo2011, Mohammadi2012, Grillo2017, IvanovPubl, Karlovets2021Vortex}. The most common technique to generate twisted electrons is to let the beam go through a phase plate \cite{Uchida2010, Schattschneider2012} or a hologram \cite{Verbeeck2010, McMorran2011, Grillo2014}, alongside with methods using surface plasmon polaritons \cite{Vanacore2019}. The states obtained with these methods can often be described as the Laguerre-Gaussian wave packets \cite{Bliokh2017, Karlovets2021Vortex}. The probability density of such states evolves in time, and a solenoid (magnetic lens) can be used to effectively control spreading of the packets, both in an electron microscope \cite{Schattschneider2014, Schachinger2015} and in a particle accelerator \cite{Reiser}.

\begin{figure}
\centering
    \includegraphics[width = 0.48\textwidth]{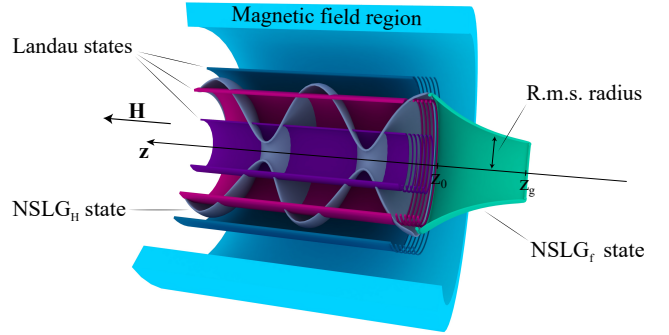}
    \caption{A free electron passing the boundary between vacuum and a solenoid (magnetic lens). Here $z_{\text{g}}$ and $z_0$ are the coordinates of the electron source and the boundary, respectively.}
    \label{fig:Lens3D}
\end{figure}

Within the hard-edge approximation, a thick magnetic lens can be described as a semi-infinite magnetic field. In a real-life experiment (see Fig. \ref{fig:Lens3D}), a free electron first propagates in vacuum towards the solenoid while spreading, then enters the lens, and continues its propagation inside it. Common description of the transmission of an electron from the field-free space to the solenoid relies on evaluating the dynamics of the observables via the Heisenberg equation of motion, and so no assumptions regarding the electron state are needed. However, far from the boundary the electron state is conventionally thought of as a stationary Landau state \cite{Greenshields2014, Greenshields2015, Karlovets2021Vortex} that does not spread in time. There have been several approaches to extend the description of an electron in the field beyond the Landau states \cite{Bagrov2002, Bliokh2012, Gallatin2012, BagrovGitman, Silenko2021, Karlovets2021Vortex, Melkani2021}. Nonetheless, the transformation of the electron state itself during the transmission process has not yet been fully understood.

The aim of the present Letter is to point out that it is the \textit{nonstationary Laguerre-Gaussian} (NSLG) states rather than the Landau ones that provide an accurate description of an electron inside a solenoid after passing the boundary. Moreover, we show that the relevance of employing the NSLG states is not only in gaining a consistent description of the electron dynamics near the boundary, but also in predicting the oscillations of the root-mean-square (r.m.s.) radius far from it. In particular, we demonstrate that the time-averaged r.m.s. radius of the NSLG state inside the field usually significantly exceeds that of the Landau state, even far away from the boundary where the Landau state might be expected to provide an adequate description. This increase in the r.m.s. radius is somewhat analogous to broadening of the classical trajectories during the synchrotron radiation \cite{STeng}, but it occurs even when no photons are emitted.


For simplicity, we illustrate our approach with a spinless electron, the transverse energy of which stays much less than $mc^2$. In principle, spin can easily be included \cite{Silenko2021}, and our approach can be extended into the relativistic regime via the light-cone variables \cite{BagrovGitman, Karlovets2015, Ducharme2021, Jentschura2023}. Throughout this Letter $\hbar = c = 1$, the electron charge $e<0$, and the electron mass equals the inverse Compton wavelength, $m = \lambda_{\text{C}}^{-1}$.

\textbf{Landau states approach}. We introduce the magnetic lens as a semi-infinite stationary and homogeneous magnetic field $\bm{H} = H\theta(z-z_0)\bm{e}_z,\, \bm{e}_z = (0,0,1)$. The step function $\theta(z)$ reflects the hard-edge boundary of the lens located at $z_0$. Detailed comments on the whole-space wavefunction in such an external field, its continuity and on the applicability of the hard-edge approximation are given in the Supplemental Material \cite{Supplemental} (see also references \cite{Lefebvre1999, Wollnik} therein). The electron wave packet propagates rectilinearly along the $z$-axis with the mean velocity $v$. To characterize the transverse dynamics of the packet, we study dynamics of the r.m.s. transverse radius $\rho(t) = \sqrt{\expv{\rho^2}(t)}$.

The wave packet generated in free space at a time $t_{\text{g}}$ is known to spread in time according to
\begin{equation}
\label{FreeDisp1}
    \rho_{\text{f}}(t) = \rho_{\text{w}}\sqrt{1+(t-t_{\text{g}})^2/\tau_{\text{d}}^2},
\end{equation}
where $\tau_{\text{d}} = \rho_{\text{w}} / u$ is the diffraction time, $\rho_{\text{w}}$ is the beam waist, $u$ is the transverse velocity dispersion and the subscript $"\text{f}\,"$ stands for "free" \cite{Karlovets2021Vortex}. As such an electron travels from the source to the lens, it acquires a non-zero divergence rate $\rho_0' = d \rho_{\text{f}} / d t |_{t = t_0}$ and its r.m.s. radius grows by a factor of  $\rho_0/\rho_{\text{w}} = \sqrt{1+(t_0 - t_{\text{g}})^2/\tau_{\text{d}}^2}$, where $t_0 = |z_0 - z_{\text{g}}|/v$ is the moment the electron enters the lens. 

Inside the field, the system is described by the Hamiltonian
\begin{equation}
    \hat{\mathcal H} = -\frac{\lambda_{\text{C}}}{2}\Delta+\frac{\omega}{2}\hat{L}_z+\frac{\omega^2}{8\lambda_{\text{C}}}{\rho}^2 = \hat{\mathcal H}_{\perp} - \frac{\lambda_{\text{C}}}{2}\partial_z^2,
\end{equation}
$\lambda_{\text{C}} = m^{-1}$ is the Compton wavelength, and $\hat{L}_z = - i \partial / \partial \varphi$ is the \textit{canonical} OAM operator. The r.m.s. radius of the electron starts oscillating according to the Heisenberg equation of motion \cite{Greenshields2014, Karlovets2021Vortex, Baturin2022}:
\begin{equation}
\label{inside}
\begin{aligned}
     \rho^2 (t) &= \rho^2_{\text{st}} + \left( \rho_0^2 -\rho^2_{\text{st}} \right) \cos\left(\omega \tau\right)+\frac{2\rho_0 \rho'_0}{\omega} \sin\left(\omega \tau\right),\\
     &\rho^2_{\text{st}} = 2\lambda_{\text{C}}\omega^{-1}\left(2\omega^{-1}\expv{\hat{\mathcal H}_{\perp}} + \expv{\hat{L}_z} \right).
\end{aligned}
\end{equation}
Here $\omega = e H \lambda_{\text{C}}$ is the cyclotron frequency and 
$\tau = t - t_0 > 0$. The subscript in $\rho_{\text{st}}$ implies the "stationary" radius; the quantity itself is the square root of period-averaged mean square radius and is the characteristic radius around which the oscillations occur.

Both in vacuum \eqref{FreeDisp1} and inside the lens \eqref{inside} the expressions for the r.m.s. radii can be obtained without specifying the electron state. Nonetheless, the latter quantitatively affects the oscillations of the r.m.s. radius by dictating $\rho^2_{\text{st}}$ in Eq.~\eqref{inside}, as well as $\rho_{\text{w}}$ and $\tau_{\text{d}}$ in Eq.~\eqref{FreeDisp1}. Far from the boundary, electrons are usually believed to be described by the Landau states, and $\rho^2_{\text{st}}$ is commonly evaluated with the aid of the Landau wave functions as
\begin{equation}
\label{st}
    \left. \rho^2_{\text{st}} \right|_{\text{Landau}} = \rho_{\text{L}}^2 = \sigma_{\text{L}}^2 (2n + \abs{l} + 1),
\end{equation}
where $\sigma_{\text{L}} = \sqrt{2/|eH|}$ and $\rho_{\text{L}}$ is the r.m.s. radius of the Landau state with a radial quantum number $n = 0,1,2,...$ and an OAM $l = 0,\pm 1,\pm 2,...$ \cite{Greenshields2014, Karlovets2021Vortex, Baturin2022}. 

As we show hereafter, it is generally \textit{not the case} that $\rho_{\text{st}} = \rho_{\text{L}}$. Eq. \eqref{st} is satisfied only for the specific boundary values $\rho_0$ and $\rho'_0$, not governed by any physical principle. In experiment, these parameters can vary from these specific values, leading to a significant increase of $\rho_{\text{st}}$ as compared to $\rho_{\text{L}}$. That affects the main characteristics of the oscillations.

\textbf{NSLG states}.
Let us find an alternative to the Landau state that would describe a twisted electron inside a solenoid after entering it from free space with arbitrary parameters $\rho_0$ and $\rho'_0$ at the boundary. Following the seminal work of Silenko et al. \cite{Silenko2022}, we note that the transverse electron wave function admits a general form, both in vacuum ($z<z_0$) and in the magnetic field ($z>z_0$) \cite{Karlovets2021Vortex, Silenko2022}:
\begin{equation}
\label{NSLG}
\begin{aligned}
    & \Psi_{n,l}(\bm{\rho},t) = N \frac{\rho^{|l|}}{\sigma^{|l|+1}(t)} L_n^{|l|} \left(\frac{\rho^2}{\sigma^2(t)}\right) \times  \\ 
    \exp & \left[il\varphi - i\Phi_{\text{G}}(t) -  \frac{\rho^2}{2\sigma^2(t)}\left(1-i\frac{\sigma^2(t)}{\lambda_{\text{C}} R(t)}\right)\right].
\end{aligned}
\end{equation}
We refer to it as a \textit{nonstationary Laguerre-Gaussian} state. The wave function \eqref{NSLG} describes a vortex electron with an OAM $l$, and the difference between the NSLG states in free space ($\operatorname{NSLG}_{\text{f}}$) and in the magnetic field ($\operatorname{NSLG}_{\text{H}}$) is governed by the optical functions: dispersion $\sigma(t)$, radius of curvature $R(t)$, and Gouy phase $\Phi_{\text{G}}(t)$.

The r.m.s. radius of the NSLG state is
\begin{equation}
\label{RMSRad}
    \rho(t) = \sigma(t) \sqrt{2n+|l|+1}.
\end{equation}
Equations for the optical functions of the $\operatorname{NSLG}_{\text{H}}$ state follow from the Schr\"{o}dinger equation:
\begin{equation}
\label{eq:Opt}
    \begin{aligned}
        &\frac{1}{R(t)} = \frac{\sigma'(t)}{\sigma(t)},\\
        &\frac{1}{\lambda_{\text{C}}^2 R^2(t)} + \frac{1}{\lambda_{\text{C}}^2} \left[\frac{1}{R(t)}\right]' = \frac{1}{\sigma^4(t)} - \frac{1}{\sigma_{\text{L}}^4},\\
        & 
    \frac{1}{\lambda_{\text{C}}}\Phi_{\text{G}}'(t) = \frac{l}{\sigma_{\text{L}}^2} + \frac{(2n+|l|+1)}{\sigma^2(t)}.
    \end{aligned}
\end{equation}
The first equation in \eqref{eq:Opt} allows us to further use  $\sigma'(t)$ rather than $R(t)$ as a characteristic of the NSLG packet.

A special choice of the initial conditions $\sigma(t_0) = \sigma_{\text{L}}$, $\sigma'(t_0) = 0$ for the system \eqref{eq:Opt} leads to a non-spreading solution
\begin{equation}
\label{Landausol}
    \sigma(t) = \sigma_{\text{L}},\, \sigma'(t) = 0,\, \Phi_{\text{G}}(t) = \varepsilon_{\perp}t,
\end{equation}
where $\varepsilon_{\perp} = \omega (2n+|l|+l+1) / 2$ is the energy of the Landau state. These optical functions turn the state \eqref{NSLG} exactly into the Landau one with $\rho_{\text{st}}$ given by Eq. \eqref{st}.

To find the more general form of the $\operatorname{NSLG}_{\text{H}}$ state, we suggest solving the system \eqref{eq:Opt} with initial conditions for the dispersion and its derivative given by the NSLG$_{\text f}$ state at the time $t_0$ when the electron enters the lens:
\begin{equation}
\begin{aligned}
     & \sigma(t_0) = \sigma_0 = \frac{\rho_0}{\sqrt{2n+|l|+1}},\\
     & \sigma'(t_0) = \sigma'_0 = \frac{\rho'_0}{\sqrt{2n+|l|+1}}.
\end{aligned}
\end{equation}
where $\rho_0$ and $\rho'_0$ are the r.m.s. radius \eqref{FreeDisp1} and the divergence rate of the $\operatorname{NSLG}_{\text{f}}$ state generated in field-free space at the time $t_{\text{g}}$, respectively. These conditions imply the continuity of the wave function at the boundary. We abstain from writing down the Gouy phase because it does not affect the dynamics of the r.m.s. radius.

The dispersion of the $\operatorname{NSLG}_{\text{H}}$ packet then reads
\begin{equation}
\label{eq:OptSolS}
    \begin{aligned}
        & \sigma(t) = \sigma_{\text{st}} \sqrt{1 + \sqrt{1 - \left( \frac{\sigma_{\text{L}}}{\sigma_{\text{st}}} \right)^4} \sin{ \left[ s(\sigma_0, \sigma_0') \omega (t-t_0) - \theta \right] }}, \\
        & \sigma_{\text{st}}^2 = \frac{\sigma_0^2}{2} \left( 1 + \left( \frac{\sigma_{\text{L}}}{\sigma_0} \right)^4 + \left( \frac{\sigma'_0 \sigma_{\text{L}}^2}{\lambda_{\text{C}} \sigma_0} \right)^2 \right), \\
        & \theta = \arcsin{\frac{1 - (\sigma_0 / \sigma_{\text{st}})^2}{\sqrt{1 - \left( \sigma_{\text{L}} / \sigma_{\text{st}} \right)^4}}},
    \end{aligned}
\end{equation}
where we have defined
\begin{equation}
    s(\sigma_0,\sigma_0') = \begin{cases}
     \operatorname{sgn}(\sigma_0'),\  \sigma_0' \ne 0,\\
     \operatorname{sgn}(\sigma_{\text{L}}-\sigma_0),\ \sigma_0'=0,\\
     0,\ \sigma_0 = \sigma_{\text{L}} \;\text{and}\; \sigma_0' = 0.
    \end{cases}
\end{equation}
Notice that the quantum numbers $n$ and $l$ do not affect the r.m.s. radius dynamics, except for scaling the oscillations magnitude according to Eq. \eqref{RMSRad}. Thus, even if a Gaussian electron with $n = l = 0$ approaches the magnetic field, the same conditions $\sigma(t_0) = \sigma_{\text{L}}$, $\sigma'(t_0) = 0$ have to hold for the electron to turn into the Landau electron inside the field.

An $\operatorname{NSLG}_{\text{H}}$ state may be thought of as a complex superposition of a large number of the Landau states \cite{SizykhArxiv}. However, it is still possible to obtain single Landau state in a solenoid. This can be achieved by 
refocusing the electron beam so that its waist is at the boundary 
and adjusting the beam size and the field strength, ensuring $\sigma_0 = \sigma_{\text{L}}$, which can be somewhat challenging to realize in experiment.

One can give the following qualitative explanation for the oscillations of the r.m.s. radius according to Eq. \eqref{eq:OptSolS} involving two processes. First, the wave packet shrinks because of the Lorentz force acting on each Bohmian trajectory. Second, the wave packet spreads similar to an $\operatorname{NSLG}_{\text{f}}$ in Eq. \eqref{FreeDisp1}. While the wave packet is wide, the radial Lorentz force, caused by the azimuthal velocity component in the longitudinal magnetic field, dominates leading to the decrease in the r.m.s. radius. As the radius becomes small enough, the quantum-mechanical spreading of the wave packet takes over and leads to expansion instead of shrinking. 

The mean energy of the $\operatorname{NSLG}_{\text{H}}$ state 
\begin{equation}
\label{eq:NSLGEn}
    \expv{E_{\perp}} = \frac{\omega}{2} (2n + \abs{l} + 1) \frac{\sigma_{\text{st}}^2}{\sigma_{\text{L}}^2} + \frac{\omega}{2} l \geq \varepsilon_{\perp}.
\end{equation}
is almost always greater than the energy of the Landau state because 
$\sigma_{\text{st}}^2 / \sigma_{\text{L}}^2 \geq 1$. 
Resulting energy excess can be attributed to the intrinsic motion of the wave packet due to the r.m.s. radius oscillations. This motion can be interpreted as quantum betatron oscillations \cite{STeng}. Such $\operatorname{NSLG}_{\text{H}}$ state's "breathing" is also reflected in a larger scale of the stationary radius $\rho_{\text{st}}^2$, when evaluated with the $\operatorname{NSLG}_{\text{H}}$ state,
\begin{equation}
\label{st2}
\left.\rho_{\text{st}}^2\right|_{\operatorname{NSLG}_{\text{H}}} = \frac{1}{T_{\text{c}}} \int\limits_0^{T_{\text{c}}} \rho^2(t) dt = \rho_{\text{L}}^2 \frac{\sigma_{\text{st}}^2}{\sigma_{\text{L}}^2} \geq \rho_{\text{L}}^2,
\end{equation}
where $T_{\text{c}} = 2 \pi m / |e H|$ is the cyclotron period.

\begin{table*}
\centering
\begin{tabular}{ | c | c | c | c | c | c | c | c | c | c | c | } 
  \hline
  Setup & $E_{\parallel}$ & $v$ & $H$ & $\rho_{\text{L}}$ & $d$ & $z_{\text{R}}$ & $\rho_0$ & $d \rho / dz |_{z = z_0}$ &  $\xi_1$ & $\xi_2$ \\ 
  \hline
  SEM & 1 keV & $0.06 c$ & 1 T & 72.6 nm & 16.3 cm & 16.3 cm & 2.82 $\mu$m & 8.7 pm/$\mu$m & 0.026 & $6.7 \times 10^{-4}$ \\ 
  \hline
  TEM & 200 keV & $0.70 c$ & 1.9 T & 52.7 nm & 10 cm & 179 cm & 2 $\mu$m & 62 pm/mm & 0.026 & $3.9 \times 10^{-5}$ \\ 
  \hline
  Medical linac & 1 MeV & $0.94 c$ & 0.1 T & 0.23 $\mu$m & 10 cm & 243 cm & 2 $\mu$m & 0.34 nm/cm & 0.115 & $5.5 \times 10^{-4}$ \\ 
  \hline
  Linac & 1 GeV & $c$ & 0.01 T & 0.72 $\mu$m & 100 cm & 258 cm & 2.14 $\mu$m & 0.28 $\mu$m/m & 0.339 & 0.045 \\ 
  \hline
\end{tabular}
\caption{Experimental scenarios for observing the oscillations of the r.m.s. radius $\rho(z)$. We take $\sigma_{\text{w}} = 1 \mu$m, $n = 0$, $l = 3$. The parameters $\xi_1 = \sigma_{\text{L}} / \sigma_0$ and $\xi_2 = \sigma'_0 \sigma_{\text{L}}^2 / (\lambda_{\text{C}} \sigma_0)$ reflect the discrepancy between the $\operatorname{NSLG}_{\text{H}}$ state and the Landau one, the latter being reproduced when $\xi_1 = 1$ and $\xi_2 = 0$.}
\label{tab:Params}
\end{table*}

\begin{figure*}
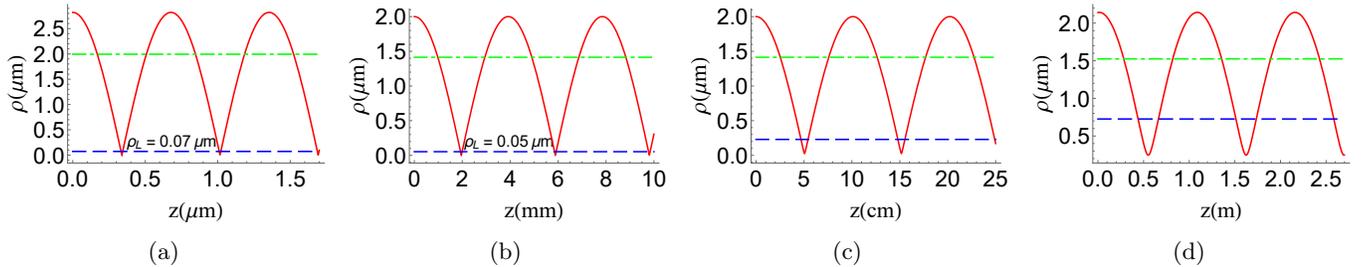

\begin{subfigure}{0.24\textwidth}
    \includegraphics[width=\textwidth]{Experimental_proposal_9_SEM.pdf}
    \caption{}
    \label{fig:ExpProp_a}
\end{subfigure}
\hfill
\begin{subfigure}{0.24\textwidth}
    \includegraphics[width=\textwidth]{Experimental_proposal_9_TEM.pdf}
    \caption{}
    \label{fig:ExpProp_b}
\end{subfigure}
 \hfill
\begin{subfigure}{0.24\textwidth}
    \includegraphics[width=\textwidth]{Experimental_proposal_9_Medical_linac.pdf}
    \caption{}
    \label{fig:ExpProp_c}
\end{subfigure}
\hfill
\begin{subfigure}{0.24\textwidth}
    \includegraphics[width=\textwidth]{Experimental_proposal_9_High_En_linac.pdf}
    \caption{}
    \label{fig:ExpProp_d}
\end{subfigure}
\caption{Oscillations of the r.m.s. radius of the $\operatorname{NSLG}_{\text{H}}$ wave packet in a magnetic field (solid red line), $\rho_{\text{st}} = (1 / T_{\text{c}}) \int_0^{T_{\text{c}}} \rho^2(t) dt$ (dot-dashed green line) and $\rho_{\text{L}}$(dashed blue line). The~parameters are listed in Table \ref{tab:Params}. (a) SEM, (b) TEM, (c) medical linac, (d) conventional linac.}
\label{fig:ExpProp}
\end{figure*}

To underline the distinction between the $\operatorname{NSLG}_{\text{H}}$ state and the Landau one we introduce
\begin{equation}
    \xi_1 = \frac{\sigma_{\text{L}}}{\sigma_0}, \quad \xi_2 = \frac{\abs{\sigma'_0} \sigma_{\text{L}}}{\lambda_{\text{C}}}.
\end{equation}
When $\xi_1 = 1$ and $\xi_2 = 0$, $\rho_{\text{st}}^2$ of Eq.~\eqref{st2} coincides with that of Eq.~\eqref{st}, obtained with the Landau state. However, in this case Eq. \eqref{inside} degenerates into $\rho^2(t) = \rho_{\text{L}}^2$ and no oscillations occur at all. 
From Eq. \eqref{st2} it follows that $\left.\rho_{\text{st}}^2\right|_{\operatorname{NSLG}_{\text{H}}} \gg \rho_{\text{L}}^2$ when either $\xi_1 \gg 1$, $\xi_1 \ll 1$ or $\xi_2 \gg 1$.

To illustrate the current approach, we compare our results with the dynamics of twisted electrons investigated experimentally in Refs.~\cite{Schattschneider2014, Schachinger2015}. The authors obtained a free electron state that, after refocusing, enters a region of a quasi-uniform magnetic field and shrinks in size while propagating inside it. During the time when the size of the electron wave packet inside the solenoid stays comparable to $\rho_{\text{L}}$, the electron is thought of as a Landau state. However, the process of the electron crossing the boundary between vacuum and magnetic field and further propagating inside might better be interpreted in the NSLG states formalism. One can reproduce the obtained behaviour of the r.m.s. radius inside the lens (see Fig.~\ref{fig:ExpProp_b} in \cite{Schattschneider2014}) using Eqs.~\eqref{eq:OptSolS}~and~\eqref{RMSRad} and the parameters $n = 0$, $\abs{l} = 1$, $\sigma_0 = 4.77 \times 10^{-2} \mu$m, $\sigma'_0 = -3.1 \times 10^{-4}$. Thus, we argue that what was observed in \cite{Schattschneider2014} may be a part of the oscillations predicted by the NSLG states approach. For the discussed experimental setup we estimate that $\xi_1 = 0.76$ and $\xi_2 = 29.21 \gg 1$ leading to
\begin{equation}
    \left.\rho_{\text{st}}^2\right|_{\operatorname{NSLG}_{\text{H}}} = 20.7\, \rho_{\text{L}} \gg \rho_{\text{L}}.
\end{equation}
Hence, typically, $\rho_{\text{st}}$ significantly exceeds $\rho_{\text{L}}$ and for oscillations to occur around the latter \textit{very specific} parameters of the incoming electron packet must align.

\textbf{Experimental feasibility}. To observe the oscillations of the r.m.s. radius described by the $\operatorname{NSLG}_{\text{H}}$ state, 
we propose several experimental scenarios for the parameters and energy scales typical for different setups: a scanning electron microscope (SEM), a transmission electron microscope (TEM), a low-energy linear accelerator (for instance, for medical applications), and a conventional linac. 
Additional solenoids must be adjusted to these electron sources to provide regions of approximately constant and homogeneous magnetic field. In an experiment, the distribution of a twisted electron probability density in the transverse plane can be  measured consecutively at various distances $z$ along the solenoid axis with, for instance, a CCD camera. Subsequently, $\sqrt{\expv{\rho^2}}$, obtained as a function of $z$, can be expressed in terms of $t = z/v$ and compared to our predictions.

The parameters for different setups are presented in Table \ref{tab:Params} and the corresponding oscillations of the r.m.s. radii are depicted in Fig. \ref{fig:ExpProp}. We take $\sigma_{\text{w}} = 1\, \mu$m (a characteristic scale \cite{Verbeeck2010, Uchida2010, McMorran2011} for the devices generating twisted electrons) and consider quantum numbers $n = 0$, $l = 3$, that results in $\rho_{\text{w}} = 2 \mu$m. A different choice of the quantum numbers would lead to simple rescaling of the r.m.s. radius according to Eq. \eqref{RMSRad}.

In Table \ref{tab:Params} we also use the longitudinal energy $E_{\parallel}$, magnetic field strength $H$, and the distance between the source of twisted electrons and the magnetic field $d$ that are typical for the proposed experimental scenarios \cite{Wiedemann}. For a SEM, we take a particular value $d = 16.3$~cm for calculation convenience, though any distance of the order of several cm is appropriate. 
We adjust the magnetic field strength in order to observe several oscillation periods at realistic distances for each setup. For instance, for a linac we take $H = 0.01$ T in order to observe oscillations at several meters. If needed, one may increase the field strength to proportionally decrease the observation distance. On the other hand, SEMs and TEMs usually have magnetic fields of the order of $1$~T and their observation distances are somewhat limited by their design.

For the $\operatorname{NSLG}_{\text{f}}$ state with $\sigma_{\text{w}} = 1\  \mu$m the diffraction time is $\tau_{\text{d}} = ~\sigma_{\text{w}}^2 / \lambda_{\text{C}} = 8.6$ ns, and the Rayleigh length, $z_{\text{R}} = v \tau_{\text{d}}$, scales with the electron energy. For example, in the second row of Table~\ref{tab:Params} the Rayleigh length for TEM, $z_{\text{R}} = 179$ cm, is much greater than the distance between the source and the solenoid, $d = 10$ cm. This leads to the r.m.s. radius at the boundary $\rho_0 \approx 2 \mu$m being almost the same as that at the electron source $\rho_{\text{w}} = 2\ \mu$m. 
The divergence rate $d \rho / dz |_{z = z_0} = \rho'_0 / v$ reflects the change in the r.m.s. radius with the distance travelled by the electron along the field near the boundary. For the proposed scenarios, $\xi_2$, the dimensionless analogue of the divergence rate, shows that the divergence rate is low and does not affect the dynamics of the electron in solenoids.

Notice the sharp wedge-like pattern of the r.m.s. radius oscillations in the bottom parts of Figs. \ref{fig:ExpProp_a} - \ref{fig:ExpProp_c}. It illustrates the influence of the parameters $\xi_1$ and $\xi_2$ on the electron behavior inside the field. Deviations of $\xi_1$ from 1 and $\xi_2$ from 0 in all the entries of Table \ref{tab:Params} emphasize the distinction between the $\operatorname{NSLG}_{\text{H}}$ state and the Landau one. For SEM, TEM and medical linac $\rho_{\text{st}}$ (dot-dashed green line in Fig. \ref{fig:ExpProp}) is \textit{almost an order of magnitude greater} than $\rho_{\text{L}}$ (dashed blue line). On the other hand, for linac (Fig. \ref{fig:ExpProp_d}) the parameters $\xi_1$ and $\xi_2$ do not differ as much from 1 and 0, correspondingly, and $\rho_{st}$ is just twice larger than $\rho_{\text{L}}$.

\textbf{Conclusion}. We have put forward an approach to the problem of transmission of a free twisted electron through a sharp boundary between a solenoid and vacuum based on the description in terms of NSLG states. This formalism enables the smooth transition of a free $\operatorname{NSLG}_{\text{f}}$ state to a single $\operatorname{NSLG}_{\text{H}}$ mode inside the field. Transformation of a free Laguerre-Gaussian electron inside the lens into the $\operatorname{NSLG}_{\text{H}}$ state leads to oscillations of the r.m.s. radius. These oscillations have usually been expected to occur around the value predicted by the stationary Landau state. Somewhat counter-intuitively, the time-averaged value of the r.m.s. radius can generally be much larger (up to several orders of magnitude) than $\rho_{\text{L}}$. For instance, for typical TEM parameters $H = 1.9$ T, $\sigma_0 = 47.7$ nm, $\sigma'_0 = -3.1 \times 10^{-4}$ from Ref. \cite{Schattschneider2014}, $\rho_{\text{st}}$ is $20$ times larger than the one predicted by the Landau states. The opposite case is $\sigma_0 \simeq \sigma_{\text{L}}$ and $\sigma'_0 \ll \lambda_{\text{C}} / \sigma_{\text{L}}$. For such parameters, the $\operatorname{NSLG}_{\text{H}}$ states resemble the Landau ones and the oscillations occur around $\rho_{\text{L}}$ with a low magnitude. 

Although there is evidence that the $\operatorname{NSLG}_{\text{H}}$ states more adequately describe quantum dynamics of the vortex electrons inside a magnetic lens, further experimental as well as theoretical scrutiny is required. For example, going beyond the hard-edge approximation, and considering an off-axis entering of an electron into a solenoid would make the proposed approach more realistic. We have proposed several scenarios that have the potential to observe the r.m.s. radius oscillations with the parameters typical for the setups ranging from SEMs to linear accelerators.

\begin{acknowledgments}
We are grateful to N. Sheremet, V.~Ivanov and S. Baturin for offering their opinion on the draft. The work is funded by Russian Science Foundation and St. Petersburg Science Foundation, project num. 22-22-20062, \url{https://www.rscf.ru/project/22-22-20062/}.
\end{acknowledgments}

\bibliographystyle{unsrt}

\begin{thebibliography}{}

\end{thebibliography}


\begin{thebibliography}{36}

\bibitem{Bliokh2017}
K.Y. Bliokh, I.P. Ivanov, G.~Guzzinati, L.~Clark, R.~{Van Boxem},
  A.~B{\'e}ch{\'e}, R.~Juchtmans, M.A. Alonso, P.~Schattschneider, F.~Nori, and
  J.~Verbeeck.
\newblock Theory and applications of free-electron vortex states.
\newblock {\em Physics Reports}, 690:1--70, 2017.

\bibitem{Lloyd2017}
S.~Lloyd, M.~Babiker, G.~Thirunavukkarasu, and J.~Yuan.
\newblock Electron vortices: Beams with orbital angular momentum.
\newblock {\em Rev. Mod. Phys.}, 89:035004, Aug 2017.

\bibitem{Verbeeck2010}
J.~Verbeeck, H.~Tian, and Peter Schattschneider.
\newblock Production and application of electron vortex beams.
\newblock {\em Nature Lett.}, 467:301--304, Sep 2010.

\bibitem{Idrobo2011}
J.C. Idrobo and S.J. Pennycook.
\newblock Vortex beams for atomic resolution dichroism.
\newblock {\em Microscopy}, 60:295--300, Oct 2011.

\bibitem{Mohammadi2012}
Z.~Mohammadi, C.P. Van~Vlack, S.~Hughes, J.~Bornemann, and R.~Gordon.
\newblock Vortex electron energy loss spectroscopy for near-field mapping of
  magnetic plasmons.
\newblock {\em Optics Express}, 20:15024--15034, 2012.

\bibitem{Grillo2017}
Vincenzo Grillo, et al.
\newblock Observation of nanoscale magnetic fields using twisted electron
  beams.
\newblock {\em Nature Comm.}, 8:689, Sep 2017.

\bibitem{IvanovPubl}
Igor~P. Ivanov.
\newblock Promises and challenges of high-energy vortex states collisions.
\newblock {\em Progress in Particle and Nuclear Physics}, page 103987, 2022.

\bibitem{Karlovets2021Vortex}
Dmitry Karlovets.
\newblock Vortex particles in axially symmetric fields and applications of the
  quantum {Busch} theorem.
\newblock {\em New Journal of Physics}, 23(3):033048, mar 2021.

\bibitem{Uchida2010}
M.~Uchida and A.~Tonomura.
\newblock Generation of electron beams carrying orbital angular momentum.
\newblock {\em Nature}, 464:737--739, Apr 2010.

\bibitem{Schattschneider2012}
P.~Schattschneider, M.~Stöger-Pollach, and J.~Verbeeck.
\newblock Novel vortex generator and mode converter for electron beams.
\newblock {\em Phys. Rev. Lett.}, 109:084801, Aug 2012.

\bibitem{McMorran2011}
Benjamin~J. McMorran, Amit Agrawal, Ian~M. Anderson, Andrew~A. Herzing,
  Henri~J. Lezec, Jabez~J. McClelland, and John Unguris.
\newblock Electron vortex beams with high quanta of orbital angular momentum.
\newblock {\em Science}, 331(6014):192--195, 2011.

\bibitem{Grillo2014}
V.~Grillo, E.~Karimi, G.C. Gazzadi, S.~Frabboni, M.R. Dennis, and R.W. Boyd.
\newblock Generation of nondiffracting electron bessel beams.
\newblock {\em Phys. Rev. X}, 4:011013, Jan 2014.

\bibitem{Vanacore2019}
G.~Vanacore, G.~Berruto, I.~Madan, et~al.
\newblock Ultrafast generation and control of an electron vortex beam via chiral plasmonic near fields.
\newblock {\em Nat. Mater.}, 18:573--579, May 2019.

\bibitem{Schattschneider2014}
P.~Schattschneider, Th. Schachinger, M.~St\"{o}ger-Pollach, S.~L\"{o}ffler,
  Steiger-Thirsfeld A., Bliokh K., and F.~Nori.
\newblock Imaging the dynamics of free-electron {Landau }states.
\newblock {\em Nature Comm.}, 5:4586, Aug 2014.

\bibitem{Schachinger2015}
T.~Schachinger, S.~Löffler, Stöger-Pollach M., and P.~Schattschneider.
\newblock Peculiar rotation of electron vortex beams.
\newblock {\em Ultramicroscopy}, 158:17--25, Nov 2015.

\bibitem{Reiser}
M.~Reiser.
\newblock {\em Theory and Design of Charged Particle Beams}.
\newblock Wiley, New York, 2008.

\bibitem{Greenshields2014}
C.R. Greenshields, R.L. Stamps, S.~Franke-Arnold, and S.M. Barnett.
\newblock Is the angular momentum of an electron conserved in a uniform
  magnetic field?
\newblock {\em Phys. Rev. Lett.}, 113:240404, Dec 2014.

\bibitem{Greenshields2015}
C.R. Greenshields, S.~Franke-Arnold, and R.L. Stamps.
\newblock Parallel axis theorem for free-space electron wavefunctions.
\newblock {\em New J. Phys.}, 17:093015, 2015.

\bibitem{Bagrov2002}
V.G. Bagrov, M.C. Baldiotti, D.M. Gitman, and I.V. Shirokov.
\newblock New solutions of relativistic wave equations in magnetic fields and
  longitudinal fields.
\newblock {\em J. Math. Phys}, 43:2284--2305, Jan 2002.

\bibitem{Bliokh2012}
Konstantin~Y. Bliokh, Peter Schattschneider, Jo~Verbeeck, and Franco Nori.
\newblock {Electron Vortex Beams in a Magnetic Field: A New Twist on Landau
  Levels and Aharonov-Bohm States}.
\newblock {\em Phys. Rev. X}, 2:041011, Nov 2012.

\bibitem{Gallatin2012}
Gregg~M. Gallatin and Ben McMorran.
\newblock Propagation of vortex electron wave functions in a magnetic field.
\newblock {\em Phys. Rev. A}, 86:012701, Jul 2012.

\bibitem{BagrovGitman}
D.~Gitman V.~Bagrov.
\newblock {\em The Dirac equation and its solutions}.
\newblock Berlin [a. o.] : de Gruyter, 2014.

\bibitem{Silenko2021}
Zou Liping, Zhang Pengming, and Alexander Silenko.
\newblock General quantum-mechanical solution for twisted electrons in a
  uniform magnetic field.
\newblock {\em Phys. Rev. A}, 103:L010201, Jan 2021.

\bibitem{Melkani2021}
Abhijeet Melkani and S.~J. van Enk.
\newblock Electron vortex beams in nonuniform magnetic fields.
\newblock {\em Phys. Rev. Research}, 3:033060, Jul 2021.

\bibitem{STeng}
A.A. Sokolov and I.M. Ternov.
\newblock {\em Radiation from Relativistic Electrons}.
\newblock American Inst. of Physics, 1986.

\bibitem{Karlovets2015}
Dmitry Karlovets.
\newblock {Gaussian and Airy} wave packets of massive particles with orbital
  angular momentum.
\newblock {\em Phys. Rev. A}, 91:013847, Jan 2015.

\bibitem{Ducharme2021}
Robert Ducharme, Irismar da~Paz, and Armen Hayrapetyan.
\newblock {Fractional Angular Momenta, Gouy and Berry Phases in Relativistic
  Bateman-Hillion-Gaussian Beams of Electrons}.
\newblock {\em Phys. Rev. Lett.}, 126:134803, Apr 2021.

\bibitem{Jentschura2023}
Ulrich Jentschura.
\newblock Algebraic approach to relativistic landau levels in the symmetric
  gauge.
\newblock {\em \eprint{arXiv:2306.01155 [hep-ph]}}, Jun 2023.

\bibitem{Supplemental}
The comments on the whole-space wavefunction are in Sec. I of Supplemental Material. In Sec. II there we demonstrate the continuity of the wavefunction implied. Sec. III is devoted to the applicability of the hard-edge approximation.

\bibitem{Lefebvre1999}
R.~Lefebvre.
\newblock {\em Continuity conditions for a time-dependent wavefunction}.
\newblock {\em THEOCHEM}, 493, 1-3: 117-123, Dec 1999.

\bibitem{Wollnik}
H.~Wollnik.
\newblock {\em Optics of Charged Particles}.
\newblock Academic Press Limited, London, 1987.


\bibitem{Baturin2022}
S.~Baturin, D.~Grosman, G.~Sizykh, and D.~Karlovets.
\newblock Evolution of an accelerated charged vortex particle in an
  inhomogeneous magnetic lens.
\newblock {\em Phys. Rev. A}, 106:042211, Oct 2022.

\bibitem{Silenko2022}
Zou Liping, Zhang Pengming, and Alexander Silenko.
\newblock Production of twisted particles in magnetic fields.
\newblock {\em \eprint{arXiv:2207.14105 [quant-ph]}}, Jul 2022.

\bibitem{SizykhArxiv}
G.K. Sizykh, A.D. Chaikovskaia, D.V. Grosman, I.I.~Pavlov, D.V. Karlovets.
\newblock Nonstationary Laguerre-Gaussian states vs Landau ones: choose your fighter.
\newblock {\em \eprint{arXiv:2309.15899 [quant-ph]}}, Sep 2023.

\bibitem{Wiedemann}
H.~Wiedemann.
\newblock {\em Particle Accelerator Physics}.
\newblock Springer Cham, 2015.

\end{thebibliography}

\end{document}


\title{Supplemental Material to "Transmission of vortex electrons through a solenoid"}

\author{G.\,K.~Sizykh}

\author{A.\,D.~Chaikovskaia}

\author{D.\,V.~Grosman}

\author{I.\,I.~Pavlov}

\author{D.\,V.~Karlovets}

{
\let\clearpage\relax
\maketitle
}
\section{On the continuity of the wave function}

In this section, we demonstrate the continuity of the approximate whole-space, or \textit{full}, wavefunction proposed in the paper. his wavefunction is constructed from two components: the wavefunction of a Laguerre-Gaussian electron in free space and in a magnetic field. We analyze the scenario where an electron, characterized by a localized wave packet, transitions from free space into a semi-infinite solenoid. For finite-length solenoids, Finite-length solenoids, this transition may be accounted by introducing the second boundary.

We are solving the Schr\"{o}dinger equation
\begin{equation}
\label{1}
    i \frac{\partial \Psi}{\partial t} = \hat{\mathcal H} \Psi
\end{equation}
with the following Hamiltonian
\begin{equation}
\label{FullSpaceHam}
    \hat{\mathcal H} = \frac{(\hat{\bm{p}} - e \bm{A} \theta(z - z_0))^2}{2m}, \quad \bm{A} = \frac{H}{2} \left\{ -y, x, 0 \right\}.
\end{equation}
The Hamiltonian can also be written as
\begin{equation}
    \hat{\mathcal H} = \hat{\mathcal H}_0 + \theta(z - z_0) \hat{\mathcal V}, \quad \hat{\mathcal H}_0 = \frac{\hat{\bm{p}}^2}{2m}, \quad \hat{\mathcal V} = - \frac{e \bm{A} \hat{\bm{p}}}{m} + \frac{(e \bm{A})^2}{2 m}.
\end{equation}

The $\theta$-function here represents the sharp boundary located at $z_0$, separating free space from the solenoid. The applicability of the \textit{hard-edge} approximation is addressed in Sec. \ref{Hard-EdgeApplicability}. Initially, we examine Eq. \eqref{1} separately for $z<z_0$ and $z>z_0$. In both regions, there is a set of nonstationary solutions of the Laguerre-Gaussian shape. The difference between the free space and magnetic states is accounted for by the optical functions, which parameterize the Laguerre-Gaussian envelopes. In our analysis, it is crucial to take a well-localized longitudinal part of the whole-space wavefunction, enabling us to use the connection between the wave packet centroid's position and time. Utilizing solutions defined distinctly for $z < z_0$ and $z > z_0$, we then construct the wave function across the entire space as follows:
\begin{equation}
\label{ansatz}
    \Psi^{\text{full}}(\bm{r},t) = \Psi_{n,l}^{\text{f}}(\bm{\rho},t)\Psi_{\parallel}(z,t) \theta(t_0-t) + \Psi_{n,l}^{\text{H}}(\bm{\rho},t)\Psi_{\parallel}(z,t)\theta(t-t_0).
\end{equation}
Here $t_0$ is the time the wave packet's centroid $\langle z \rangle$ reaches the boundary at $z_0$; $\theta(t - t_0)$ and $\theta(t_0-t) = 1 - \theta(t - t_0)$ reflect the transition process and $\Psi^{(1, 2)}(\bm{r},t) = \Psi^{\text{f,H}}(\bm{\rho},t) \Psi_{\parallel}(z,t)$ are the free-space and magnetic-field solutions accordingly:
\begin{equation}
    \begin{aligned}
        & i \frac{\partial \Psi^{(1)} (\bm{r},t)}{\partial t} = \hat{\mathcal H}_0 \Psi^{(1)}, \\
        & i \frac{\partial \Psi^{(2)} (\bm{r},t)}{\partial t} = \left( \hat{\mathcal H}_0 + \hat{\mathcal V} \right) \Psi^{(2)}.
    \end{aligned}
\end{equation}
Note the absence of $\theta$-functions in these equations.

Since the longitudinal magnetic field solely influences the transverse dynamics of the wave packet, we assume identical longitudinal wave functions in both regions without any loss of generality. Despite their somewhat exotic nature, the time-dependent $\theta$-functions play a vital role in ensuring the continuity of the whole-space wavefunction at all times. Notice also that with the aid of the longitudinal motion law,
\begin{equation}
\label{EquationsOfMotion}
    \langle z \rangle = z_0 + v(t-t_0) \Rightarrow \langle z \rangle - z_0 = v(t-t_0),
\end{equation}
Eq. \eqref{ansatz} can be written in a form with a more clear physical sense:
\begin{equation}
\label{ansatzAvgz}
    \Psi^{\text{full}}(\bm{r},t) = \Psi_{n,l}^{\text{f}}(\bm{\rho},t)\Psi_{\parallel}(z,t)\theta(z_0-\langle z \rangle) + \Psi_{n,l}^{\text{H}}(\bm{\rho},t)\Psi_{\parallel}(z,t)\theta(\langle z \rangle-z_0).
\end{equation}
Here, $\theta$-functions switch between the solutions based on the longitudinal position of the wave packet's centroid.

Before we delve into analyzing the continuity of the proposed ansatz, we first want to set the stage by describing the physical scenario we aim to depict and outlining the approximations we employ. We are considering a longitudinally well-localized wave packet transitioning from free space to the solenoid. At a specific moment in time, its probability density is centered around $z_0$. We imply that at this time, when the free electron enters the magnetic field, the free-space solution continuously transforms into the magnetic one. In doing so, we treat the wave packet as if it enters the solenoid as a whole, similar to a point-like particle. However, in reality, there exists a time interval (or a distance in space) during which one tail of the wave packet is already inside the solenoid while the other remains in free space. The complete transfer of the wave packet from one region to another is a gradual process, during which the state may undergo transitional effects. We justify neglecting these transitional processes by confirming that the diffraction time of the free-space solution and the cyclotron period are significantly greater than the characteristic time required for the wave packet to traverse the transition region. Consequently, the wave packet remains almost unchanged during this transition. A detailed analysis supporting this justification is provided in the subsequent section.

To demonstrate the continuity of the whole-space wavefunction, we require the wavefunction given by \eqref{ansatz} or Eq. \eqref{ansatzAvgz} to be an \textit{approximate} solution of Eq. \eqref{1}, meaning
\begin{equation}
\label{ApproximateSolution}
    i \frac{\partial \Psi^{\text{full}}}{\partial t} - \hat{\mathcal H} \Psi^{\text{full}} = i \left( \Psi^{(2)} - \Psi^{(1)} \right) \delta(t - t_0) + \theta(z_0 - z) \theta(t - t_0) \hat{\mathcal V} \Psi^{(2)} - \theta(z - z_0) \theta(t_0 - t) \hat{\mathcal V} \Psi^{(1)} \rightarrow 0.
\end{equation}
From the first term we get the condition
\begin{equation}
\label{ContinuityCondition}
    \Psi^{\text{I}}_{n,l}(\bm{\rho},t_0) = \Psi^{\text{II}}_{n,l}(\bm{\rho},t_0).
\end{equation}
The two other terms are proportional to $\theta(z_0 - z) \theta( \langle z \rangle - z_0)$ and $\theta(z - z_0) \theta(z_0 - \langle z \rangle)$, respectively. These terms vanish for a point-like particle, rendering $\Psi^{\text{full}}$ an exact solution of Eq. \eqref{1}. However, for a smeared, even well-localized, wave packet, the solution $\Psi^{\text{full}}$ becomes approximate. The effect of spacial spreading of the wave packet is discussed in the subsequent section.

Therefore, the wavefunction \eqref{ansatz}, which describes an electron transitioning from free space into a solenoid, as implied in the paper, serves an approximate solution to the corresponding Schr\"{o}dinger equation, provided that the condition \eqref{ContinuityCondition} is met and the wave packet is well-localized in the longitudinal direction.

Further we demonstrate that even being an approximate solution to Eq. \eqref{1}, the wavefunction \eqref{ansatz} is strictly continuous:
\begin{equation}
\label{cont}
    \partial_{t}\rho^{\text{full}}(\bm{r},t) + \nabla \cdot \bm{j}^{\text{full}}(\bm{r},t) = 0,
\end{equation}
where the probability density and current are given by expressions
\begin{equation}
    \rho^{\text{full}}(\bm{r},t) = |\Psi^{\text{full}}(\bm{r},t)|^2, \quad \bm{j}^{\text{full}}(\bm{r},t) = \lambda_\text{C}\text{Im}\left[\Psi^{\text{full}*}(\bm{r},t)\nabla\Psi^{\text{full}}(\bm{r},t)\right] - \lambda_\text{C} e \bm{A} \theta(z - z_0) \rho^{\text{full}}(\bm{r},t).
\end{equation}
Indeed, introducing probability densities and currents for the solutions in free space and in the solenoid,
\begin{equation}
    \begin{aligned}
        & \rho^{(1)} = |\Psi^{(1)}|^2, \quad \bm{j}^{(1)} = \lambda_\text{C} \text{Im} \left[\Psi^{(1)*} \nabla \Psi^{(1)} \right] - \lambda_\text{C} e \bm{A} \theta(z - z_0) \rho^{(1)}, \\
        & \rho^{(2)} = |\Psi^{(2)}|^2, \quad \bm{j}^{(2)} = \lambda_\text{C} \text{Im} \left[\Psi^{(2)*} \nabla \Psi^{(2)} \right] - \lambda_\text{C} e \bm{A} \theta(z - z_0) \rho^{(2)},
    \end{aligned}
\end{equation}
and taking $\theta(t - t_0) \theta(t_0 - t) \equiv 0$ and $\theta(z - z_0) = 1 - \theta(z_0 - z)$ into account one finds
\begin{equation}
    \begin{aligned}
        & \partial_t \rho^{\text{full}}(\bm{r},t) - \nabla \cdot \bm{j}^{\text{full}}(\bm{r},t) = \left[ \partial_{t} \rho^{(1)} \theta(t_0 - t) + \partial_{t} \rho^{(2)} \theta(t - t_0) + (\rho^{(2)} - \rho^{(1)}) \delta(t - t_0) \right] \\
        - & \left[ \nabla \cdot \bm{j}^{(1)} \theta(t_0 - t) + \nabla \cdot \bm{j}^{(2)} \theta(t - t_0) + \lambda_\text{C} e \theta(t_0 - t) \nabla \cdot (\bm{A} \rho^{(1)} \theta(z_0 - z)) + \lambda_\text{C} e \theta(t - t_0) \nabla \cdot (\bm{A} \rho^{(2)} \theta(z_0 - z)) \right].
    \end{aligned}
\end{equation}

Bearing in mind continuity equations
\begin{equation}
    \begin{aligned}
        & \partial_{t} \rho^{(1)} + \nabla \cdot \bm{j}^{(1)} = 0, \\
        & \partial_{t} \rho^{(2)} + \nabla \cdot \bm{j}^{(2)} = 0,
    \end{aligned}
\end{equation}
we arrive at
\begin{equation}
    \begin{aligned}
        & \partial_t \rho^{\text{full}}(\bm{r},t) - \nabla \cdot \bm{j}^{\text{full}}(\bm{r},t) = (\rho^{(2)} - \rho^{(1)}) \delta(t - t_0) \\
        - & \left[ \lambda_\text{C} e \theta(t_0 - t) \nabla \cdot (\bm{A} \rho^{(1)} \theta(z_0 - z)) + \lambda_\text{C} e \theta(t - t_0) \nabla \cdot (\bm{A} \rho^{(2)} \theta(z_0 - z)) \right].
    \end{aligned}
\end{equation}
The last two terms in the r.h.s. of this equation are equal to zero. Indeed,
\begin{equation}
\label{SupportiveEquation}
    \nabla \cdot (\bm{A} \rho^{(i)} \theta(z_0 - z)) = (\nabla \cdot \bm{A}) \rho^{(i)} \theta(z_0 - z) + \bm{A} \cdot (\nabla \rho^{(i)}) \theta(z_0 - z) + \rho^{(i)} \bm{A} \cdot (\nabla \theta(z_0 - z)) = 0, \quad i = \{ 1, 2 \}.
\end{equation}
As $\bm{A}$ is given by the second equation in Eq. \eqref{FullSpaceHam}, it can be checked directly that $(\nabla \cdot \bm{A}) = 0$. In the second term in the r.h.s. of Eq. \eqref{SupportiveEquation} we can write $\bm{A}$ in cylindrical coordinates as $\bm{A} = H \bm{e}_{\varphi}$; additionally, due to cylindrical symmetry of the problem, $(\nabla \rho^{(i)}) \cdot \bm{e}_{\varphi}$, and the second term is zero. The third term vanishes due to $(\nabla \theta(z_0 - z)) \propto \bm{e}_z$ and $\bm{A} \cdot \bm{e}_z = 0$.

Finally, we are left with
\begin{equation}
    \partial_t \rho^{\text{full}}(\bm{r},t) - \nabla \cdot \bm{j}^{\text{full}}(\bm{r},t) = (\rho^{(2)} - \rho^{(1)}) \delta(t - t_0),
\end{equation}
which is zero due to the condition \eqref{ContinuityCondition} we impose on the wavefunction. Thus, the ansatz we use in the paper is a continuous function at all times.

\section{On the instantaneous transfer to the solenoid}

In this section we discuss temporal and spatial regions, in which the approximate whole-space wavefunction we imply in the paper does not accurately describe the transfer of an electron from free space to a solenoid.

Let us assume the Gaussian form for the longitudinal wave packet:
\begin{equation}
\label{WFLongitudinal}
\begin{aligned}
    \Psi_{\parallel} =  \int\limits_{-\infty}^{\infty} d p_z &\exp{i \left( p_z (z - z_0) - \frac{p_z^2}{2m} (t - t_0) \right)} \exp{- \frac{(p_z - p)^2}{2 \sigma_p^2}}  = \sqrt{\frac{2 \sqrt{2} \pi}{\sigma_z^2 + \tilde{t}^2/(m \sigma_z)^2 }} \exp{- \frac{\left( \tilde{z} - v \tilde{t} \right)^2}{2 \left( \sigma_z^2 + \tilde{t}^2/(m \sigma_z)^2 \right)}}\\\
    &\times \exp{i \left( p \tilde{z} - \frac{p^2}{2m} \tilde{t} \right)}  \exp{ -i \arctan{\frac{\tilde{t}}{m \sigma_z^2}}} \exp{i \frac{\left( \tilde{z} - v \tilde{t} \right)^2}{2 \left( \sigma_z^2 + \tilde{t}^2/(m \sigma_z)^2  \right)} \frac{\tilde{t}}{m \sigma_z^2}}.
\end{aligned}
\end{equation}
Here $\tilde{z} = z - z_0, \tilde{t} = t - t_0$, $z_0$ is the position of the boundary dividing the vacuum and magnetic field regions, $t_0$ is the moment of time when the electron probability density is centered around $z_0$, $\sigma_p$ and $\sigma_z = \sigma_p^{-1}$ are characteristic longitudinal sizes of the wave packet at the boundary in momentum and coordinate spaces, respectively, and $v = p / m$ is the average longitudinal velocity, where $p$ is the average longitudinal momentum.

\subsection{Time constraints}

The distinctive time it takes the wave packet to enter the solenoid is the time it needs to travel the distance $\sigma_z$, which is equal to $\sigma_z / v$. To neglect any transitional effects arising from the finite time of entering the solenoid, we require the characteristic timescales, during which the transverse parts of the free space and magnetic solutions undergo changes, to be much greater than $\sigma_z/v$. The time evolution of the $\operatorname{NSLG}_{\text{f}}$ state is dictated by the diffraction time $\tau_{\text{d}}$, while that of the $\operatorname{NSLG}_{\text{H}}$ state is determined by the cyclotron period $T_{\text{c}} = 2 \pi / \omega$. Thus, the criterion for the validity of the approximation of instantaneous transfer to the solenoid is
\begin{equation}
\label{TimeConditions}
    \tau_{\text{d}}, T_{\text{c}} \gg \sigma_{z}/v.
\end{equation}
Let us now make some quantitative estimates. $T_{\text{c}} \approx 36 / H[\text{T}]$ ps, $\tau_{\text{d}} = \rho_{\text{w}} / u = \rho_{\text{w}}^2 / [(2n + |l| + 1) \lambda_{\text{C}}] \approx 8.6 \rho_{\text{w}}^2[\text{$\mu$m}] / (2n + |l| + 1)$ ns, and $\sigma_z / v = (\sigma_z / c) (c / v) \approx 3.3 (c / v) \sigma_z[\text{nm}]$ as. As an illustrative example, we take typical parameters of a vortex electron in a scanning electron microscope: $H = 1$ T, $\rho_{\text{w}} = 2 \mu$m, $n = 0$, $|l| = 3$, $v / c = 0.06$, $\sigma_z \sim 1$ nm. Then both conditions are satisfied with a margin: $\tau_{\text{d}} \approx 8.6$ ns $\gg \sigma_z / v \approx 55$ as; $T_{\text{c}} \approx 36$ ps $\gg \sigma_z / v \approx 55$ as.

To violate the condition $T_{\text{c}} \gg \sigma_z / v$, one either has to take incredibly slow electrons, $v \sim 10^{-7} c$, or extremely wide wave packets, $\sigma_z \sim 1$ mm, or weaken the magnetic field by 6 orders of magnitude. To breach the condition $\tau_{\text{d}} \gg \sigma_z / v$, smaller values of $\rho_{\text{w}}$ can by taken alongside with higher values of the OAM. For example, $\rho_{\text{w}} = 0.03 \mu$m, $n = 0$, $|l| = 99$, $\sigma_z = 30$ nm, and $v = 0.001 c$ lead to $\tau_{\text{d}} \approx 77$ ps $\sim \sigma_z / v = 100$ fs.

Thus, the conditions \eqref{TimeConditions} are generally satisfied, and our approach is applicable in nearly any sensible scenario.

\subsection{Space constraints}

As previously indicated with the aid of Eq. \eqref{ApproximateSolution}, the ansatz \eqref{ansatz} provides an approximate solution to the Schrödinger equation \eqref{1} for a particle spread out in space. By employing the longitudinal wavefunction of Gaussian type \eqref{WFLongitudinal}, we determine a characteristic attenuation length. In this subsection, we ascertain this length and identify the space regions where the two last terms in Eq. \eqref{ApproximateSolution} vanish, ensuring that the implied approximate solution accurately describes a real electron.

The dependence of the probability density on the longitudinal coordinate, according to Eq. \eqref{WFLongitudinal}, is given by the exponent
\begin{equation}
    \rho_{\parallel} \propto \exp{- \frac{\left( \tilde{z} - v \tilde{t} \right)^2}{\left( \sigma_z^2 + \tilde{t}^2/(m \sigma_z)^2 \right)}}.
\end{equation}
To facilitate analysis, we introduce dimensionless quantities: Compton wavelength $\Lambda_{\text{C}} = \lambda_{\text{C}} / \sigma_z = 1 / (m \sigma_z)$, longitudinal coordinate $\zeta = \tilde{z} / \sigma_z$, and time $\tau = v \tilde{t} / \sigma_z$. The latter denotes the position of the wave packet's centroid $\langle z \rangle$ according to the equation of motion \eqref{EquationsOfMotion}. Note that $\Lambda_{\text{C}} \ll 1$ in any realistic scenario. The $\theta$-functions in the last two terms of Eq. \eqref{ApproximateSolution} correspond to either the wavefunction tail being outside the field, $z < z_0$, at times $t > t_0$ after the wavefunction's centroid entered the field, or vice verse, the wavefunction tail being inside the field, $z > z_0$, at times $t < t_0$ before the wavefunction's centroid entered the field. In both scenarios, the signs of $\tilde{z}$ and $\tilde{t}$ are opposite, causing the exponent to transform into
\begin{equation}
    \rho_{\parallel} \propto \exp{- \frac{\left( \zeta + \tau \right)^2}{1 + \tau^2 \Lambda_{\text{C}}^2}}.
\end{equation}
Now we analyze the behavior of this probability density in different parameter regions.
\begin{enumerate}
    \item Little times, $\tau \ll \zeta$, $\tau \Lambda_{\text{C}} \ll 1$.
        \begin{equation}
        \rho_{\parallel} \propto \exp{-\zeta^2}.
        \end{equation}
    In this case, the approximation holds true for $\zeta \gtrsim 1$, i.e., outside the region $(z_0 - \sigma_z, z_0 + \sigma_z)$,. This is a tiny region near the boundary, provided the electron wave packet has the typical size $\sigma_z$ on the order of several nanometers. The larger the size of the electron, the more extensive the area near the boundary where the approximation inadequately describes the actual wavefunction.
    
    \item Large distances, $\tau \ll \zeta$, $\tau \Lambda_{\text{C}} \gg 1$.
        \begin{equation}
        \rho_{\parallel} \propto \exp{- \frac{\zeta^2}{\tau^2 \Lambda_{\text{C}}^2}}.
        \end{equation}
    In this regime, $\zeta \gtrsim \tau \Lambda_{\text{C}}$ must hold, which is always the case as $\Lambda_{\text{C}} \ll 1$ and $\tau \ll \zeta$. Consequently, for sufficiently large times where $\tau \Lambda_{\text{C}} \gg 1$, the wavefunction we imply is a good approximation to the exact solution to the Schr\"{o}dinger equation \eqref{1} far away from the boundary. It's noteworthy that in this domain, $\zeta \gg \Lambda_{\text{C}}^{-1}$.
    
    \item Little times, little distances, $\tau \gg \zeta$, $\tau \Lambda_{\text{C}} \ll 1$.
        \begin{equation}
        \rho_{\parallel} \propto \exp{-\tau^2}.
        \end{equation}
    In this very narrow region near the boundary, $\zeta \ll \Lambda_{\text{C}}^{-1}$, $\tau \gtrsim 1$ must be fulfilled.
    
    \item Large times, $\tau \gg \zeta$, $\tau \Lambda_{\text{C}} \gg 1$.
        \begin{equation}
        \rho_{\parallel} \propto \exp{- \frac{1}{\Lambda_{\text{C}}^2}}.
        \end{equation}
        In this region, the probability density is always small and the approximation holds provided that $\tau \gg \zeta$ and $\tau \Lambda_{\text{C}} \gg 1$.
\end{enumerate}

In conclusion, the approximate wavefunction \eqref{ansatz} implied in the paper effectively describes the electron's transition to the solenoid across all times and space regions, except for a narrow zone near the boundary with a characteristic width $\sigma_z$.

\section{On the applicability of the hard-edge approximation}
\label{Hard-EdgeApplicability}

Here, we justify the use of the hard-edge approximation, that is, replacement of the field of a real solenoid $\tilde{\bm{H}}$ by $\bm{H} = H \theta(z - z_0) \bm{e}_z$. The approximation neglects: 1) a smooth change in the field's $z-$component; 2) the nonzero radial component of the field. For further discussion, it is useful to recall the Bohmian interpretation of quantum mechanics and consider an electron wave packet as a quantum analogue of a classical particle beam.

In accelerator physics it is shown that the smoothness of the field's longitudinal component simply leads to replacing the actual length of the solenoid $d_0$ with
\begin{equation}
    d = \int \limits_{-\infty}^{\infty} \frac{\tilde{H}_z^2(z)}{H^2} dz,
\end{equation}
where $\tilde{H}_z(z)$ is the $z-$component of the field of a real solenoid at the $z-$axis and $H$ is its maximal value \cite{Reiser, Wollnik}. This can be accounted for by applying boundary conditions at the point $z_0 - (d - d_0) / 2$, rather than at $z_0$.

Now we will consider the influence of the radial field component in the SEM scenario. Generalizing to relativistic particles is straightforward but technical.

From Maxwell's equations, it follows that
\begin{equation}
\label{FieldsConnection}
    \tilde{H}_{\rho} (\rho, z) = -\frac{1}{2} \rho \tilde{H}'_z(z)
\end{equation}
in the first order in the transverse coordinate. This leads to the longitudinal $F_z = e v_{\varphi} \tilde{H_{\rho}}$ and radial $F_{\varphi} = e v \tilde{H_{\rho}}$ Lorentz forces, where $v_{\varphi}$ and $v$ are the azimuthal and longitudinal velocities, respectively. The longitudinal force induces acceleration (deceleration), and neglecting the field's radial component is justified if the change in the longitudinal energy of the electron is negligible. The azimuthal force leads to a change in the azimuthal kinetic velocity that has no effect on the canonical OAM.

Assuming $v$ stays almost constant and using $z \approx v t$ and $d / dt = v d / dz$, the system of equations of motion turns into
\begin{equation}
\label{LorentzEqs}
    \frac{d v_{\varphi}}{d z} v = \frac{e v \tilde{H}_{\rho}}{m}, \hspace{30pt} \frac{d v}{d z} = \frac{e v_{\varphi} \tilde{H}_{\rho}}{m v}.
\end{equation}
The first equation, with Eq. \eqref{FieldsConnection} taken into account, leads to $v_{\varphi}(z) = v_{\varphi}^0 - e \rho \tilde{H}_z / (2 m)$
with $v_{\varphi}^0 = v_{\varphi}(- \infty)$. Then, the second equation can be written as
\begin{equation}
    \frac{d v^2}{d z} = \left( \frac{e \rho}{2m} \right)^2 \frac{d \tilde{H}_z^2}{d z} - \frac{e \rho v_{\varphi}^0}{m} \frac{d \tilde{H}_z}{dz}.
\end{equation}
With conditions $\tilde{H}_z(-\infty) = 0$, $\tilde{H}_z(\infty) = H$, this induces the total change of the longitudinal energy:
\begin{equation}
\label{TotalChange}
    \Delta E_{\parallel} = \frac{m \Delta v^2}{2} = \frac{m}{2} \left[ \left( \frac{e \rho \tilde{H}}{2m} \right)^2 - \frac{e \rho \tilde{H}}{m} v_{\varphi}^0 \right].
\end{equation}

Let us now estimate the change in the longitudinal energy in the SEM scenario from the first row in Table 1 and Fig. 2(a) in the main text, for which $\rho = \rho_{\text{st}} \approx 2 \mu$m, $H = 1$ T, and $E_{\parallel} = 1$ keV. The value $v_{\varphi}^0$ can be estimated via the mean transverse energy given by Eq. (12) in the main text: $v_{\varphi}^0 \leq \sqrt{2 m \langle E_\perp \rangle}$. For $n \sim l \sim 1$ we have $v_{\varphi}^0 \sim \sqrt{2 m\expv{E_\perp} } \sim \lambda_{\text{C}} \sigma_{\text{st}} / \sigma_{\text{L}}^2 \approx 10^{-3}$ and, according to Eq. \eqref{TotalChange}, $\Delta E_{\parallel} \approx - 0.21$ eV or $\Delta E_{\parallel}  \approx 0.38$ eV, depending on the sign of the second term in Eq. \eqref{TotalChange}. Thus, in this scenario, the change in $E_{\parallel}$ is on the level of $0.01 \%$. Due to $v_{\varphi}^0 \propto \sqrt{|l|}$ at large $|l|$, even at the highest achievable values of $|l|$ up-to-date \cite{Mafakheri2017}, which are on the order of several thousand, fringe fields of real solenoids can be ignored.

For other experimental scenarios discussed, $\Delta E_{\parallel}$ is even smaller than $0.01 \%$ because of the decrease in the field strength at almost the same $\rho$ (see Table 1 in the main text).

\bibliographystyle{unsrt}